\documentclass[aps,twocolumn]{revtex4}
\usepackage{color}
\usepackage{epsfig}
\usepackage{graphics}
\begin{document}

\title{Spectroscopic Studies of Fractal Aggregates of Silver
  Nanospheres Undergoing Local Restructuring}

\author{Sergei V. Karpov}
\affiliation{L.V. Kirensky Institute of Physics, Russian Academy of
  Sciences, Siberian Branch, Krasnoyarsk 660036, Russia}

\author{Valeriy S. Gerasimov and Ivan L. Isaev}
\affiliation{Department of Physics and Engineering, Krasnoyarsk State
  Technical University, Krasnoyarsk 660028, Russia}

\author{Vadim A. Markel}
\affiliation{Departments of Radiology and Bioengineering,
  University of Pennsylvania, Philadelphia, PA 19104}

\begin{abstract}
  We present an experimental spectroscopic study of large random
  colloidal aggregates of silver nanoparticles undergoing local
  restructuring. We argue that such well-known phenomena as strong
  fluctuation of local electromagnetic fields, appearance of ``hot
  spots'' and enhancement of nonlinear optical responses depend on the
  local structure on the scales of several nanosphere diameters,
  rather that the large-scale fractal geometry of the sample.
\end{abstract}

\date{\today}
\maketitle

Physical properties of surface plasmons (SPs) in disordered
nanosystems have been the subject of intense studies in the past
decade~\cite{kreibig_book_95,stockman_00_2,shalaev_book_00,roldugin_03_2}.
The two objects that attracted most attention are disordered 
two-component
composites~\cite{bergman_80_1,aspnes_82_1,stockman_99_1},
including two-dimensional percolation films~\cite{seal_03_1}, and
aggregates of nanometer-sized noble metal spheres formed in colloidal
solutions (colloidal
aggregates)~\cite{butenko_90_1}.  The latter are subject
of this paper. Optical and, more generally, electromagnetic effects
that were discovered in colloidal aggregates include giant enhancement
of nonlinear-optical
responses~\cite{butenko_90_1,stockman_92_1},
inhomogeneous localization of electromagnetic
eigenmodes~\cite{stockman_96_1,stockman_97_1}, and optical
memory~\cite{karpov_88_1,kim_99_1}.

Perhaps, the most fundamental physical feature of electromagnetic
interaction in colloid aggregates is inhomogeneous
broadening~\cite{markel_91_1}. That is, different electromagnetic
modes that can be excited in such aggregates (collective SP
excitations) are resonant at different wavelengths which form a
continuous band extending from the optical to the far-IR spectral
region~\cite{danilova_93_1,markel_96_1}. In contrast, a single
isolated silver nanosphere has a well defined narrow resonance
centered at approximately $\lambda=400{\rm nm}$ (in
hydrosols)~\cite{kreibig_book_95}. The spectral shifts of the
collective SP excitations are explained by the electromagnetic
interaction of nanospheres. We view such excitations as collective
even though not all nanospheres may effectively participate in a
particular excitation modes. In fact, it was shown, that at any given
electromagnetic frequency, there exist SP excitation which are
delocalized over the whole sample, as well as excitations localized on
a few neighboring nanospheres (hot
spots)~\cite{stockman_96_1,stockman_97_1}. It was also shown, both
experimentally and in simulations, that the locations of these hot
spots are very sensitive to the electromagnetic frequency and
polarization of the incident wave~\cite{markel_99_3}.

The spatial properties of the collective SP excitations in random
colloidal aggregates have been studied in great detail (see
Refs.~\cite{kreibig_book_95,stockman_00_2,shalaev_book_00,roldugin_03_2}
and references therein).  However, little is known about the relation
between the sample geometry and spatial properties of electromagnetic
eigenmodes.  Typically, the eigenmodes are obtained as solutions to
the electromagnetic interaction problem.  More specifically, an
infinite matrix representing the electromagnetic interaction operator
is truncated and diagonalized numerically~\cite{markel_04_3}.  Each
element of this matrix is completely defined by the sample geometry.
In principle, the same is also true for the eigenmodes. However, the
mathematical dependence between elements of a large matrix and its
eigenvectors can be very complex and, in the general case, not easily
analyzable. Some approximate theories that directly relate the spatial
characteristics of the sample and the electromagnetic field excited in
the sample were based on the first Born~\cite{martin_87_1} and
mean-field~\cite{berry_86_1} approximations, on few-body interaction
approximation (binary~\cite{markel_91_1} or
binary-ternary~\cite{stockman_97_1} approximations), and various
phenomenological scaling
laws~\cite{markel_91_1,stockman_95_1,shalaev_92_1}.
However, the first Born and the mean-field approximations are not
applicable to resonant excitation of collective SPs.  The few-body
approximations and the scaling laws proved to be very useful for
qualitative theoretical description at the early stages of research,
but increasingly more realistic simulations revealed that these
approaches do not provide quantitative results.

We have recently argued that the locations of ``hot spots'' in random
fractal aggregates are strongly correlated with the local anisotropy
factor~\cite{karpov_05_1} which quantifies the deviation of the local
environment of a given nanosphere in an aggregate from the spherical
symmetry. We have shown in simulations that sites with high local
anisotropy are likely to coincide with the ``hot spots''. This concept
allows one to make a qualitative prediction about location of the
``hot spots'' in a large aggregate without actually solving the
electromagnetic problem. In this paper we present experimental
evidence of this conjecture.

Direct measurements of local fields near the surface of aggregates
deposited on a flat substrate is possible with near-field scanning
optical microscopy~\cite{markel_99_3}, while the
geometrical structure can be probed with atomic-force microscopy or
electron microscopy. However, these methods only yield two-dimensional
images and are not suitable for investigating three-dimensional
structure of sample or of the field.  Therefore, direct measurement of
the local anisotropy factor in three-dimensional samples is a
difficult experimental task. (We note that this can be achieved, in
principle, by solving the inverse scattering problem in the near
field~\cite{carney_04_1}.)  However, it is possible to observe the
influence of local restructuring on the electromagnetic properties of
a large aggregate indirectly by studying IR absorption of large
nanoaggregates as they undergo local restructuring. The enhancement of
IR absorption in an aggregate relative to an isolated nanosphere is an
important effect explained by the inhomogeneous broadening, or
spectral shifts of resonance eigenmodes from the Fr\"{o}hlich
frequency of an isolated nanosphere into the IR region. The
inhomogeneous spectral broadening is a direct consequence of strong
resonance interaction of different nanospheres in an aggregate and re
To this end, we have studied extinction spectra of silver fractal
aggregates embedded in a polymer matrix under uniform contraction.  We
have studied how the long wavelength wings of the absorption spectra
evolve due to the contraction.

Experimental samples were prepared as follows. First, silver hydrosol
was prepared by reduction of ${\rm AgNO_3}$ by ${\rm NaBH_4}$ in water
solution~\cite{butenko_90_1} (electrolyte concentration
$2.5\cdot10^{-3}$~M). After this chemical reaction, the hydrosol is a
colloidal solution of silver nanospheres $10{\rm nm}$ to $40{\rm nm}$
in diameter that undergo random Brownian motion and can stick on
contact and form large fractal aggregates. The aggregation process was
accelerated by irradiation of the hydrosol with the natural light for
$4-10{\rm min}$~\cite{karpov_02_2}. The overall size of aggregates was
of the order of $1\mu{\rm m}$ or larger as follows from TEM images.
Next, the hydrosol containing aggregates of silver nanospheres was
added into a water solution of gelatin. The volume fraction of silver
in the resultant solution was $\sim 10^{-6}$. Next, the prepared
solution was allowed to gelate.  As a result, the metal nanoparticles
became rigidly connected to the polymer matrix of the gel and could
not move freely.  We have prepared rectangular gelatin samples with
initial dimensions of $1\times 5 \times 10{\rm cm}^3$.  At the next
stage, the gel underwent gradual dehydration and the sample volume was
reduced by the factor of $\approx 10$. This corresponds to linear
contraction by the factor of $\approx 2.15$.

Electron micrographs of $Ag$ nanoaggregates in the hydrosol and in
thin slices of the gel (several hundreds nm thick) taken after the
dehydration are shown in Fig.~1. It can be seen that, as the gelatin
matrix undergoes contraction, the local geometrical structure of the
aggregates changes. Neighboring nanospheres tend to form dense blobs
which are spherically symmetric on average. Inside such blob, the
local environment of a nanosphere is more similar to that in chaotic
dense packing.  Therefore, we expect that the local asymmetry factor
is reduced in the restructured aggregates.\\

\psfig{file=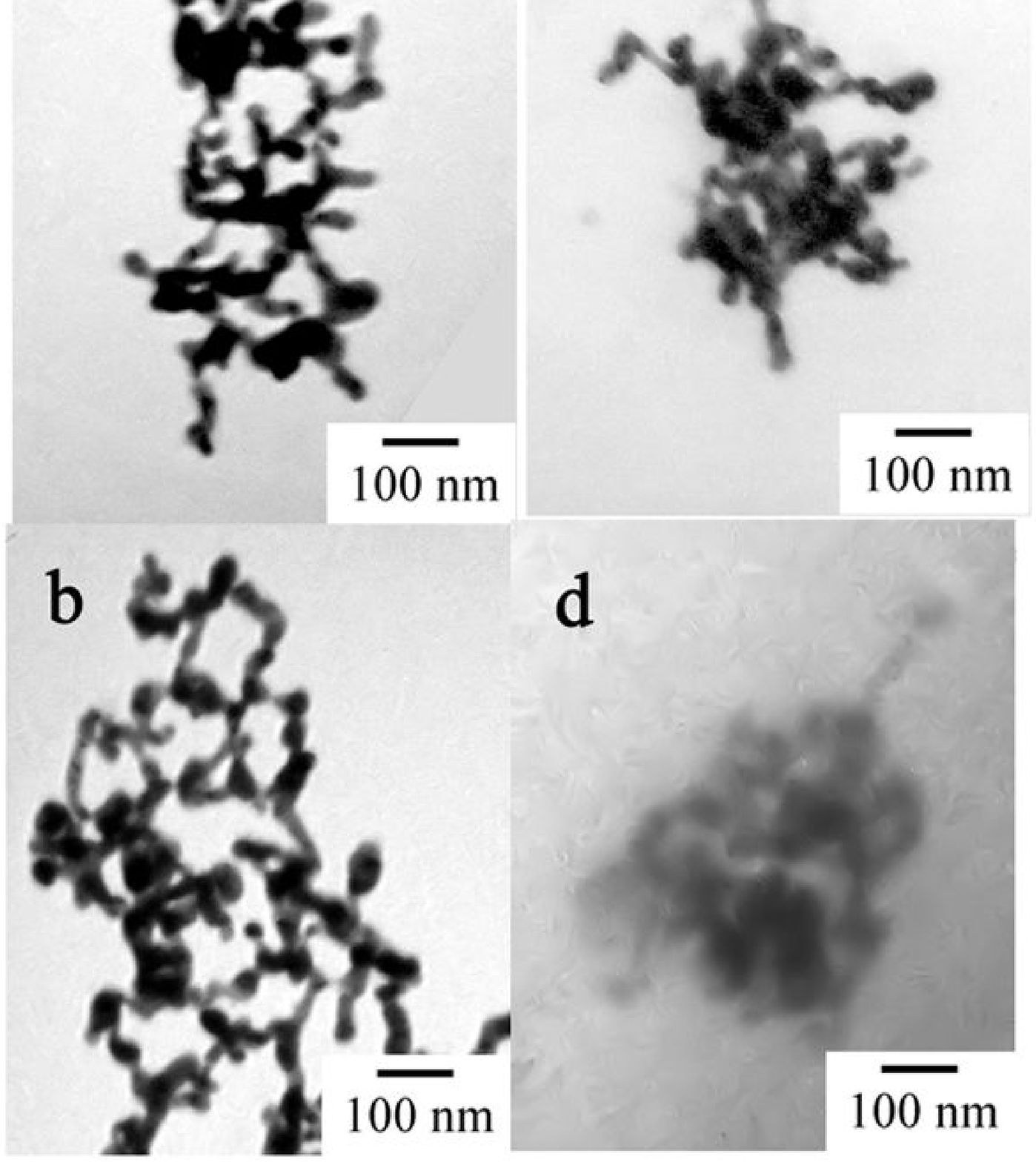,width=8.2cm}\vspace*{-1.5cm}
{\small Fig.~1. Comparative electron micrographs of typical silver
  nanoaggregates before their embedding into the gelatin matrix (a, b)
  and in slices of the gel in the final stage of the dehydratation (c,
  d). The slice thickness is approximately 200-300 nm (c) and 500-700
  nm (d). Low resolution in image (d) is explained by the relatively
  large slice thickness. Nanoaggregates shown in the left and right
  columns are not the same.} \\

We have also simulated the restructuring process on a computer.
Initially, a random two-dimensional (2D) aggregate with $N=700$
nanospheres and fractal dimension $D\approx 1.46$ and a
three-dimensional (3D) aggregate with $N=2000$ nanospheres and fractal
dimention $D\approx 1.83$ were generated using the method described in
Ref.~\cite{markel_04_3}. Then the coordinates of each nanosphere were
repeatedly multiplied by the factor $0.95$, while the sphere radii
were kept constant. The contraction resulted in geometrical
intersection of neighboring spheres. Then each sphere (in a
predetermined order) was moved to the nearest possible position as to
avoid its intersection with any other sphere in the aggregate. Next
the contraction by the factor $0.95$ was repeated, and so on, until
the overall linear contraction of the sample by the factor $\approx
2.5$ was achieved (approximately, in 50 iterations). The stages of
transformation of the aggregate are shown in Fig.~2.  The local
restructuring similar to the one seen in TEM images is clearly
visible.  However, the global geometrical structure of aggregates
remains approximately unchanged, within the limits imposed by the
changing overall size of the aggregate. Note that the the changes of
local structure in 3D aggregates are partially obscured by the
overlaps in 2D projection of a 3D object. This problem is not present
in the case of 2D aggregates, which are shown in Fig.~2 for
illustrative purpose. \\

\psfig{file=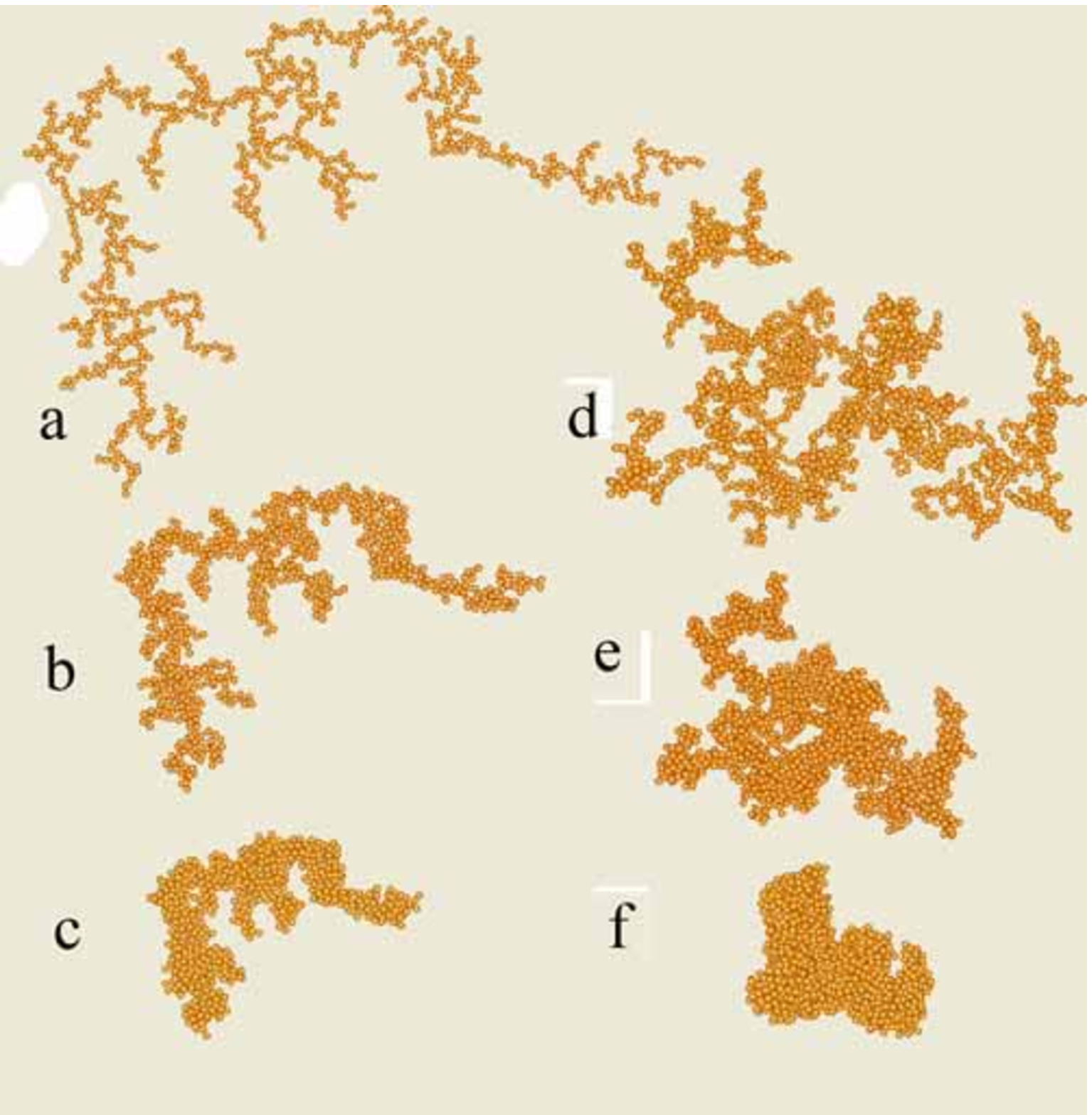,width=8.2cm}
{\small Fig.~2. Simulation of stages of transformation of 2D and 3D
  fractal aggregates embedded in a contracting gelatin matrix.
  Original 2D and 3D aggregates (a,d) after linear contraction by the
  factor $\sim 1.4$ (b,e) and $\sim 2.5$ (c,f). For the 3D aggregate,
  the average value of the local anisotropy parameter $S$ defined in
  Ref.~\cite{karpov_05_1} are 1.03 (d), 0.94 (e) and 0.74 (f).}\\

The evolution of the absorption spectra of the samples as they undergo
gradual contraction are shown in Fig.~3. Here the solid bold line
(line 1) represents the absorption spectrum of silver hydrosol before
aggregation, the solid thin line (line 2) the spectrum after
aggregation and before embedding in the gelatin matrix. Dashed lines
3-5 describe different stages of dehydration of the matrix. These
curves correspond to the absorption of silver aggregates; the
absorption of pure gelatin was subtracted from the composite samples.
It can be seen that, as the aggregates undergo restructuring due to
the matrix contraction, the long wavelength spectral wing is
suppressed. In particular, the absorption at $\lambda\approx 800{\rm
  nm}$ in the dehydrated gelatin matrix (curve 5) is reduced
approximately by the factor of $2$ compared to that in the original
gelatin matrix (curve 3) and by the factor of $3$ compared to
aggregates in solution. We note that at that wavelength, the
absorption in the dehydrated matrix is close to that of non-aggregated
nanospheres in solution. This indicates that all the effects related
to inhomogeneous broadening, such as the giant local field
fluctuations, appearance of ``hot spots'' and enhancement of nonlinear
responses are strongly suppressed in the aggregates that underwent
local restructuring. The spectral changes have been reproduced in a
series of five independent experiments. \\

\psfig{file=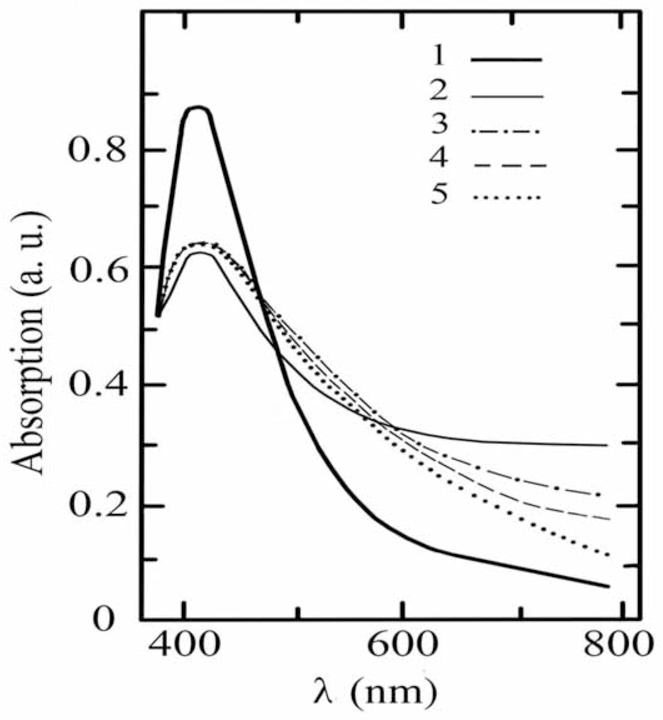,width=8.2cm,height=8.2cm}
{\small Fig.~3. Absorption spectra of silver hydrosol in
  non-aggregated stage (1) and in the final stage of aggregation (2),
  and of silver aggregates embedded in the gelatin matrix in different
  stages of dehydration of gelatin (3-5). Curves 3-5 show differential
  absorption spectra: the absorption of identical gelatin matrix
  without silver aggregates was subtracted. Absorpotion spectra are
  shown at the initial stage of freshly prepared matrix (curve 3), at
  the intermediate stage of dehydration (curve 4) and after full
  dehydration (curve 5). Curves 2-5 are normalized to common
  maximum.}\\

In summary, we have provided experimental evidence that the local
anisotropy of the environment introduced in Ref.~\cite{karpov_05_1},
rather than the large-scale geometrical structure, is the crucial
factor for fluctuation and enhancement of local fields in random
aggregates of nanospheres.

This research was supported by the Russian Foundation for Basic
Research, Grant 05-03-32642. The authors are grateful to Dr. O.P.
Podavalova for preparation of gelatin matrixes and Dr. S.M. Zharkov
for TEM images. Contact author: S.V.~Karpov (karpov@iph.krasn.ru).

\bibliographystyle{prsty} 
\bibliography{abbrevplain,master,book}

\end{document}